# Toward Model Matching for Remotely Controlled Differential Drive Robotic Vehicles

Nikolaos D. Kouvakas[1], Fotis N. Koumboulis[1],
Konstantinos G. Tzierakis[1,2], John Sigalas[1] and Anastasios Dimakakos[2]
[1] Robotics, Automatic Control and Cyber-Physical Systems Laboratory,
Department of Digital Industry Technologies, School of Science,
National and Kapodistrian University of Athens, Euripus Campus, Euboea, Greece
[2] Core Department,
National and Kapodistrian University of Athens, Euripus Campus, Euboea, Greece
`nkouvak@dind.uoa.gr, fkoumboulis@dind.uoa.gr, tzikos@uoa.gr,`
`giansiga@uoa.gr,adimakakos@core.uoa.gr`

**Abstract**

The problem of regulation of the orientation angle of a remotely controlled differential-drive mobile robot with actuator dynamics and network-induced delays is studied. Using a preinstalled two-layer nonlinear control scheme that decouples linear and angular velocities and regulates heading, a third, delay-dependent layer that achieves exact model matching from the orientation angle command to the orientation angle is introduced. The proposed outer loop controller is a delay dependent dynamic measurable output-feedback controller with dynamic proper precompensator. Parameterization yields a simple characteristic quasi-polynomial with coefficients constrained to satisfy stability for all delays up to a computable bound. Computational experiments confirm accurate tracking, fast settling and bounded internal signals and control voltages. The approach offers an analytic design alternative to AI-based tuning for delayed robotic systems.

## 1   Introduction

Differential drive robot vehicles are popular mobile robotic platforms because of their relatively simple and efficient locomotion mechanism (Cobos Torres 2013; Cobos Torres et al. 2014; Dhaouadi and Abu Hatab 2013; Drosou et al. 2024a and 2024b; Kamel and Zhang 2014; Kouvakas et al. 2022; Kouvakas et al. 2024; Martins et al 2020; Rubio et al. 2019). They offer high maneuverability in confined spaces, and so they are beneficial in indoor navigation, warehouses, and service robotics. They are also offered as testbeds in research for sophisticated control techniques and path planning.

Remote control of differential drive robot vehicles (Jung et al. 2011; Xie et al. 2020; Zeiger et al 2009) is critical when human intervention is not possible or risky, such as in hazardous environments

(e.g., heavy industry). The integration of the robot and the operator with a communication network adds instability factors such as network-induced time delays, packet losses, and bandwidth limitations. This factors would prevent the application standard software tools and analytic formulae for the design of controllers. Such factors usually decrease system stability and effectiveness, which incites the use of AI-based methodologies (Kusuma et al. 2016; Osler and Sands 2022; Sambana and Ramesh 2020).

In the present paper, the problem of regulation of the orientation angle of a remotely controlled differential-drive mobile robot in the presence of network-induced delays is studied. Here, the robotic vehicle is considered to initially be equipped with a preinstalled two-layer nonlinear scheme that decouples linear and angular velocity of the system and regulates the heading angle. Based upon the resulting closed loop system, a third, analytically determined delay-dependent layer that achieves exact model matching from the orientation angle command to the orientation angle is developed. By appropriate selection of the controller parameters a simple quasi-polynomial characteristic equation is derived. Its coefficients are constrained to guarantee stability for all delays up to a computable bound. The performance of the proposed control scheme is demonstrated through simulations. The proposed strategy presents an analytic, reproducible alternative to AI-based tuning with explicit parameterization and stability assurance that removes trial-and-error search, delivers predictable behavior, and enables the design to be systematically replicated/adapted across various platforms.

## 2 Preliminaries

In the present subsection, the nonlinear dynamics of a differential drive mobile robot will be presented. The model of the robot is produced assuming pure rolling and no lateral slip conditions as well as that the robot's active wheels are powered by suitable DC motors (Anvari 2013, Cobos Torres 2013; Cobos Torres et al. 2014; Dhaouadi and Abu Hatab 2013; Drosou et al. 2024a and 2024b; Kouvakas et al. 2022 and 2024). According to Kouvakas et al. (2024) and assuming no external disturbances or uncertainties, the nonlinear dynamic model governing its motion is of the form $\tilde{\mathbf{E}}(\mathbf{x})\mathbf{x}^{(1)} = \tilde{\mathbf{A}}\mathbf{x} + \tilde{\mathbf{B}}\mathbf{u}$ and $\mathbf{y} = \mathbf{C}\mathbf{x}$, where $\mathbf{x} = \begin{bmatrix} x_1 & x_2 & x_3 & x_4 & x_5 \end{bmatrix}^T = \begin{bmatrix} \omega_{W,l} & \omega_{W,r} & \varphi & i_{m,l} & i_{m,r} \end{bmatrix}^T$, $\mathbf{u} = \begin{bmatrix} u_1 & u_2 \end{bmatrix}^T = \begin{bmatrix} V_{m,l} & V_{m,r} \end{bmatrix}^T$, $\mathbf{y} = \begin{bmatrix} y_1 & y_2 \end{bmatrix}^T = \begin{bmatrix} v & \varphi \end{bmatrix}^T$, where $\mathbf{x}$, $\mathbf{u}$ and $\mathbf{y}$ are the state, input and performance output vectors, respectively, and $\omega_{W,l}$ and $\omega_{W,r}$ are the left and right active wheel angular velocities respectively, $\varphi$ is the vehicle's orientation angle, $i_{m,l}$ and $i_{m,r}$ are the left and right motor currents, respectively, $V_{m,l}$ and $V_{m,r}$ are the left and right motor voltage supplies, respectively, and $v$ is the linear velocity of the vehicle. The system matrices are $\tilde{\mathbf{E}}(\mathbf{x}) \in \mathbb{R}^{5\times 5}$, $\tilde{\mathbf{A}} \in \mathbb{R}^{5\times 5}$, $\tilde{\mathbf{B}} \in \mathbb{R}^{5\times 2}$ and $\mathbf{C} \in \mathbb{R}^{2\times 5}$. The analytic expressions of the system matrices in terms of the parameters of vehicle are presented in Kouvakas et al. (2024). The measurable variables are the elements of $\boldsymbol{\psi} = \begin{bmatrix} \psi_1 & \cdots & \psi_6 \end{bmatrix}^T$ where $\boldsymbol{\psi} = \mathbf{C}_{m,0}\mathbf{x} + \mathbf{C}_{m,1}\mathbf{x}^{(1)} + \mathbf{C}_{m,2}\nabla_{\tau_1}\mathbf{x} + \mathbf{C}_{m,3}\nabla_{\tau_2}\mathbf{x}^{(1)}$, where $\tau_1$ and $\tau_2$ are time varying transmission delays and $\nabla_\lambda f = f(t - \lambda(t))$. The non-zero elements of $\mathbf{C}_{m,k} = \left[ (c_{m,k})_{i,j} \right] \in \mathbb{R}^{6\times 5}$ ($k = 0, \ldots, 3$) are $(c_{m,0})_{1,1} = (c_{m,0})_{2,2} = (c_{m,1})_{3,1} = (c_{m,1})_{4,2} = (c_{m,2})_{5,3} = (c_{m,3})_{6,3} = 1$. Here it holds that $\tau_1 = \tau_2 = \tau$, where $\tau$ is a transmission delay by application of the algorithm in Koumboulis et al. (2016).

The controller proposed in Kouvakas et al. (2024), is preinstalled to the robotic vehicle, aiming toward regulation of the performance variables of the system. This preinstalled controller is of the form

$$u_1 = \frac{\lambda_{0,1}}{2a_{1,5}}w_1 + \frac{\lambda_{0,2}}{2a_{2,6}}w_2 + \frac{a_{2,5}}{2a_{2,6}}\tilde{y}_1^{(1)}\tilde{y}_2^{(1)} + \frac{a_{2,3}}{2a_{2,6}}\tilde{y}_1\tilde{y}_2^{(2)} + \frac{a_{1,3}}{2a_{1,5}}\tilde{y}_2^{(1)}\tilde{y}_2^{(2)} + \frac{a_{2,1}-\lambda_{1,2}}{2a_{2,6}}\tilde{y}_2^{(2)} + \frac{a_{1,1}-\lambda_{1,1}}{2a_{1,5}}\tilde{y}_1^{(1)} +$$
$$\frac{a_{1,4}}{2a_{1,5}}\left(\tilde{y}_2^{(1)}\right)^2 + \frac{a_{2,4}}{2a_{2,6}}\tilde{y}_1\tilde{y}_2^{(1)} + \frac{a_{2,2}-\lambda_{0,2}}{2a_{2,6}}\tilde{y}_2^{(1)} + \frac{a_{1,2}-\lambda_{0,1}}{2a_{1,5}}\tilde{y}_1, \ u_2 = \frac{\lambda_{0,1}}{2a_{1,5}}w_1 - \frac{\lambda_{0,2}}{2a_{2,6}}w_2 - \frac{a_{2,5}}{2a_{2,6}}\tilde{y}_1^{(1)}\tilde{y}_2^{(1)}$$
$$-\frac{a_{2,3}}{2a_{2,6}}\tilde{y}_1\tilde{y}_2^{(2)} + \frac{a_{1,3}}{2a_{1,5}}\tilde{y}_2^{(1)}\tilde{y}_2^{(2)} + \frac{a_{1,1}-\lambda_{1,1}}{2a_{1,5}}\tilde{y}_1^{(1)} + \frac{a_{1,4}}{2a_{1,5}}\left(\tilde{y}_2^{(1)}\right)^2 - \frac{a_{2,4}}{2a_{2,6}}\tilde{y}_1\tilde{y}_2^{(1)} + \frac{\lambda_{0,2}-a_{2,2}}{2a_{2,6}}\tilde{y}_2^{(1)} + \frac{a_{1,2}-\lambda_{0,1}}{2a_{1,5}}\tilde{y}_1,$$
$$w_2 = -\rho_1\left(\nabla_\tau y_2^{(1)} + \psi_{n,6}\right) - \rho_0\left(\nabla_\tau y_2 + \psi_{n,5}\right) + \kappa\tilde{w}_2$$

where $w_1$ and $\tilde{w}_2$ are the external commands to the linear velocity and the orientation angle, respectively, $\lambda_{i,j}$ ( $i=0,1$, $j=1,2$ ), $\rho_0$, $\rho_1$ and $\kappa$ are real free controller parameters, $\tilde{y}_1 = r_W\left(\psi_1+\psi_2\right)/2$, $\tilde{y}_1^{(1)} = r_W\left(\psi_3+\psi_4\right)/2$, $\tilde{y}_2^{(1)} = r_W\left(\psi_2-\psi_1\right)/(2b_W)$, $\tilde{y}_2^{(2)}(t) = r_W\left(\psi_2^{(1)}-\psi_1^{(1)}\right)/(2b_W)$, $r_W$ and $b_W$ are the active wheels' radius and the half distance between the hubs of the active wheels, respectively, and $a_{i,j}$ are functions of the model parameters, analytically presented in Kouvakas et al. (2024).

## 3 A Delay Dependent Controller for Model Matching

In Kouvakas et al. (2024), a metaheuristic algorithm has been employed to tune the controller parameters, because of the difficulty to calculate all gains analytically. Though analytic expressions are calculated for a set of parameters to maintain stability and delay tolerance, the remaining degrees of freedom influence transient performance and robustness in a way not quantified by closed-form expressions, especially regarding approximate model matching for the transfer function mapping $\tilde{w}_2$ to the orientation angle. Metaheuristic search is a pragmatic searching mean for the remaining parameters and obtains good performance. Clearly, an analytic technique, if feasible, would be preferred. Analytic tuning yields theoretical guarantees, interpretability, and computational efficiency.

Applying the controller in Section 2 to the nonlinear model of the robotic vehicle, it can be verified (Kouvakas et al. 2024) that the closed loop system description takes on the form

$$y_1^{(2)} + \lambda_{1,1}y_1^{(1)} + \lambda_{0,1}y_1 = \lambda_{0,1}w_1, \quad y_2^{(3)} + \lambda_{1,2}y_2^{(2)} + \lambda_{0,2}y_2^{(1)} + \lambda_{0,2}\rho_1\nabla_\tau y_2^{(1)} + \lambda_{0,2}\rho_0\nabla_\tau y_2 = \lambda_{0,2}\kappa\tilde{w}_2 \quad (1)$$

Considering the decoupled form of the closed loop system, we propose a third layer delay dependent controller, toward achieving exact model matching (Koumboulis et al. 2007; Koumboulis et al. 2009) for the transfer function mapping the external command to the orientation angle. The third layer controller is $\tilde{W}_2(s) = K_1(s,z)\Psi_5(s) + K_2(s,z)\Psi_6(s) + G(s,z)R(s)$, where $\tilde{W}_2(s) = \mathcal{L}\{\tilde{w}_2(t)\}$, $\Psi_5(s) = \mathcal{L}\{\psi_5(t)\}$, $\Psi_6(s) = \mathcal{L}\{\psi_6(t)\}$, $R(s) = \mathcal{L}\{r(t)\}$, $\mathcal{L}\{\bullet\}$ denotes the Laplace Transform of the argument signal, $r$ is the external command the orientation angle, $z = \exp(-s\tau)$, $K_1(s,z)$, $K_2(s,z)$ and $G(s,z)$ are rational functions of the Laplace transform variable $s$ whose numerator and denominator polynomial coefficients are, in general, rational functions of $z$. Choosing $\kappa = \eta_0/\eta_1$, $\lambda_{0,2} = \eta_1\mu_0$, $\lambda_{1,2} = \eta_1 + \mu_0$, $\rho_0 = \eta_0/\eta_1$ and $\rho_1 = \eta_0/\eta_1\mu_0$ (Kouvakas et al. 2024), where $\eta_0$, $\eta_1$ and $\mu_0 \in \mathbb{R}$ are arbitrary real parameters, the forced response of the second equation in (1) becomes $Y_2(s) = \kappa\lambda_{0,2}H_{w,2}(s,z)\tilde{W}_2(s)$, where $H_{y,2}(s,z) = 1/\left(p_{c,1}(s)p_{c,2}(s,z)\right)$, $p_{c,1}(s) = s + \mu_0$ and

$p_{c,2}(s,z) = s^2 + \eta_1 s + \eta_0 z$. Application of the third layer controller results in $Y_2(s) = H_c(s,z)R(s)$, where $H_c(s,z) = G(s,z)\eta_0\mu_0 / \{s(s+\eta_1)(s+\mu_0) + \eta_0[sz + (z - K_1(s,z) - sK_2(s,z))\mu_0]\}$, under the constraint $K_1(s,z) \neq [\kappa\lambda_{0,2}H_{w,2}(s,z)]^{-1} - K_2(s,z)s$. Here, $K_1(s,z)$ and $K_2(s,z)$ are selected to be static, i.e. $K_1(s,z) = k_1$ and $K_2(s,z) = k_2$ where $k_1, k_2 \in \mathbb{R}$. The closed loop transfer function mapping the orientation command to the orientation angle becomes $H_c(s,z) = G(s,z)\eta_0\mu_0 / p_a(s,z)$ where $p_a(s,z) = s^3 + (\eta_1 + \mu_0)s^2 + [\eta_1\mu_0 + z\eta_0(1 - k_2\mu_0)]s + z(1 - k_1)\eta_0\mu_0$. Clearly, the feedback elements must be chosen such that the closed loop system is stable. Let $\mu_0 = \chi_1$, $\eta_0 = \chi_3/(1 - k_2\chi_1)$, $\eta_1 = \chi_2$ and $k_1 = k_2\chi_1$, where $\chi_j$ ($j = 1,\ldots,4$) are real parameters under the constraint $k_2\chi_1 \neq 1$. The characteristic quasi-polynomial takes on the form $p_a(s,z) = (s + \chi_1)(s^2 + \chi_2 s + z\chi_3)$. Following the procedure presented in Kouvakas et al. (2024), the parameters of the quasi-polynomial are constrained to satisfy the inequalities $\chi_1, \chi_2, \chi_3 \in \mathbb{R}^+$, $\chi_3 < \chi_2^2/4$, $\chi_1 \neq (\chi_2 \pm \sqrt{\chi_2^2 - 4\chi_3})/2$.

The closed loop system characteristic quasi-polynomial is stable for all $\tau \in [0, \tau_{max})$, where $\tau_{max} = 2\tan^{-1}(T_c\sqrt{\chi_3/(1 + T_c\chi_2)}) / \sqrt{\chi_3/(1 + T_c\chi_2)}$ and $T_c = -\chi_2^{-1} + 0.5\chi_2\chi_3^{-1} + 0.5\sqrt{4\chi_2^{-2} + \chi_2^2\chi_3^{-2}}$. Regarding exact model matching problem, let $H_m(s,z)$ be the model transfer function. Choosing $G(s,z) = (1 - k_2\chi_1)\chi_1^{-1}\chi_3^{-1}p_a(s,z)H_m(s,z)$, exact model matching for the transfer function mapping the external command to the orientation angle has been achieved. Clearly, if $H_m(s,z)$ is stable, then $G(s,z)$ is also stable. For the precompensator to be proper, if $n_n$ and $n_d$ are the orders (with respect to $s$) of the numerator and denominator polynomials of $H_m(s,z)$ it must hold that $n_d \geq n_n + 3$.

## 4  Simulation Results

To illustrate the performance of the proposed controller, the vehicle model and preinstalled controller parameters, as well as the operating points in Kouvakas et al. (2024) are used. Regarding exact model matching for the orientation angle, let $H_m(s,z) = 1/[(T_{c,1}s + 1)(T_{c,2}s + 1)(T_{c,3}s + 1)]$, where $T_{c,1} = 0.04$, $T_{c,2} = 0.05$ and $T_{c,3} = 0.06$. According to Kouvakas et al. (2024), let $\chi_1 = 1.42662$, $\chi_2 = 217.2061$ and $\chi_3 = 676.2171$. Furthermore, let $k_2 = 0.1$. Clearly, the stability constraints in Section 3 are satisfied. For demonstration, the external commands are $w_1(t) = \bar{y}_1(1 + 0.1u_s(t))$ and $r(t) = \bar{y}_2(1 + 0.2u_s(t))$, where $u_s(t)$ is the unit step signal. In Figures 1 and 2, the resulting closed loop responses of the performance variables are presented. Closed-loop simulation results verify the effectiveness of the proposed controller. The output responses are the same as the desired model's responses. The ideal overlap between the nonlinear system outputs and the reference paths verifies that tracking is ensured accurately. From simulation results it can be observed that the internal dynamics and actuator-level variables also display satisfactory characteristics. The wheel angular velocities are bounded, while the motor currents and the supply voltages display transients typical of the required effort.

# 5 Conclusions

In the present paper, the problem of the regulation of the orientation angle of a remotely controlled differential-drive mobile robot in the presence of network-induced delays has been achieved. Building on a preinstalled two-layer nonlinear scheme that decouples linear and angular velocities and regulates heading, a third control layer, being a delay-dependent layer, that achieves exact model matching from the orientation angle command to the orientation angle, has been developed. The proposed outer loop controller is of dynamic measurable output-feedback form with dynamic proper precompensator. By appropriate selection of the controller parameters an elegant characteristic quasi-polynomial is derived, whose coefficients are constrained to guarantee stability for all delays up to a computable bound. Computational experiments confirm accurate tracking and fast settling, as well as appropriately bounded internal signals and control voltages. The approach offers an analytic, reproducible alternative to AI-based tuning for delayed robotic systems. Future work includes a) investigation of the performance of the proposed approach in the presence of parametric uncertainties, external disturbances and measurement noise, b) the use of adaptive or learning-based extensions to enable online controller parameter tuning under changing conditions, and c) extension of the proposed framework to multi-robot coordination / consensus problems, where synchronization and communication delays are critical.

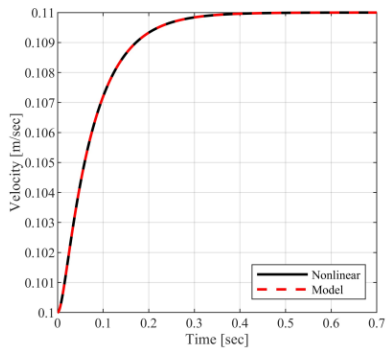
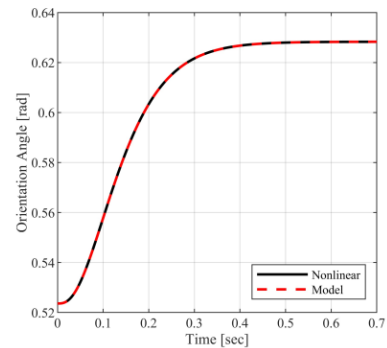

**Figure 1:** Closed loop linear velocity response.  **Figure 2:** Closed loop orientation angle response.